\documentclass[aps,pre,reprint,groupedaddress,notitlepage,showpacs]{revtex4-1}
\usepackage{graphicx}
\usepackage{amsmath}
\usepackage{amsfonts}
\usepackage{amssymb}
\usepackage{color}

\newcommand{\omegabold}{\boldsymbol\omega}

\begin{document}

\title{Statistical properties of the velocity field for the 3D hydrodynamic turbulence onset}
\thanks{dmitrij@itp.ac.ru}

\author{D.S. Agafontsev$^{(a),(b)}$, E.A. Kuznetsov$^{(b),(c),(d)}$ and A.A.
Mailybaev$^{(e)}$}
\affiliation{$^{(a)}$ P. P. Shirshov Institute of Oceanology of RAS, Moscow, Russia\\
	$^{(b)}$ Skolkovo Institute of Science and Technology, Moscow, Russia\\
	$^{(c)}$ P.N. Lebedev Physical Institute of RAS, Moscow, Russia\\
	$^{(d)}$ L.D. Landau Institute for Theoretical Physics of RAS, Moscow, Russia\\
	$^{(e)}$ Instituto Nacional de Matem\'atica Pura e Aplicada -- IMPA, Rio de Janeiro, Brazil}

\begin{abstract}
We study the statistical correlation functions for the three-dimensional hydrodynamic turbulence onset when the dynamics is dominated by the pancake-like high-vorticity structures. 
With extensive numerical simulations, we systematically examine the two-points structure functions (moments) of velocity. 
We observe formation of the power-law scaling for both the longitudinal and the transversal moments in the same interval of scales as for the energy spectrum. 
The scaling exponents for the velocity structure functions demonstrate the same key properties as for the stationary turbulence case. 
In particular, the exponents depend on the order of the moment non-trivially, indicating the intermittency and the anomalous scaling, and the longitudinal exponents turn out to be slightly larger than the transversal ones. 
When the energy spectrum has power-law scaling close to the Kolmogorov's one, the longitudinal third-order moment shows close to linear scaling with the distance, in line with the Kolmogorov's $4/5$-law despite the strong anisotropy. 
\end{abstract}

\maketitle

%----------------------------------------------------------------------------
%----------------------------------------------------------------------------

\textbf{1.} Despite great practical importance, there are only a few exact results known for turbulence theory. 
The basic result is the Kolmogorov's $4/5$-law~\cite{kolmogorov1941dissipation,landau2013fluid,frisch1999turbulence}, which for the inertial interval of scales $r$ is written as
\begin{equation}
\langle\delta v_{\parallel}^{3}\rangle = -(4/5)\,\varepsilon\,r,
\label{K45}
\end{equation}
where $\delta v_{\parallel}$ is the longitudinal variation of velocity, $\varepsilon$ is the mean energy dissipation in unit mass, and $\langle...\rangle$ denotes ensemble-averaging. 
Using dimensional analysis, Kolmogorov also found relations for the second-order structure functions, $\langle\delta v^{2}\rangle\propto \varepsilon^{2/3}r^{2/3}$, and the energy spectrum, $E_{k}\propto\varepsilon^{2/3} k^{-5/3}$. 
Kolmogorov's arguments are based on the assumptions of statistical homogeneity and isotropy of the flow and also locality of nonlinear interaction at the scales of the inertial interval. 
Then, the dynamics at these scales can be described by the Euler equations and the emergence of the Kolmogorov's relations may be expected before the viscous scales get excited~\cite{orlandipirozzoli2010,holm2002transient,cichowlas2005effective,holm2007}. 

In particular, as we demonstrated in our previous papers~\cite{agafontsev2015,agafontsev2016development}, the power-law energy spectrum with close to Kolmogorov's scaling can be observed in a fully inviscid flow when its dynamics is dominated by the pancake-like high-vorticity structures~\cite{brachet1992numerical,agafontsev2016asymptotic,agafontsev2017universal}. 
Such structures generate strongly anisotropic vorticity field in the Fourier space, concentrated in ``jets'' extended in the directions perpendicular to the pancakes. 
These jets, occupying only a small fraction of the entire spectral space, dominate in the energy spectrum, leading to formation of the power-law interval $E_{k}\propto k^{-\alpha}$ with the exponent $\alpha$ close to $5/3$ and expanding with time to smaller scales. 
Moreover, the power-law scaling extends significantly longer if the emerging jets align close to the same direction, increasing the anisotropy of the flow. 

In this paper we continue these studies and present numerical evidence that, despite the strong anisotropy, the $4/5$-law may also be satisfied before the viscous scales get excited. 
With numerical simulations of the three-dimensional Euler equations, we examine the two-points structure functions (moments) of velocity. 
We observe formation of the power-law scaling $[M_{\parallel}^{(n)}(r)]^{1/n}\propto r^{\xi_{n}}$ and $[M_{\perp}^{(n)}(r)]^{1/n}\propto r^{\zeta_{n}}$ for both the longitudinal and the transversal moments in the same interval of scales as for the energy spectrum $E_{k}$. 
The scaling exponents $\xi_{n}$ and $\zeta_{n}$ demonstrate the same key properties as for the developed (stationary) turbulence case: they decrease with the order $n$ of the moment, indicating the intermittency and the anomalous scaling, and the longitudinal exponents turn out to be slightly larger than the transversal ones. 
Analyzing simulations for different initial conditions, we observe an approximate relation $\xi_{3}\simeq\alpha/5$, so that when the power-law scaling of the energy spectrum is close to the Kolmogorov's one, the longitudinal third-order moment shows close to linear scaling with the distance, compatible with the Kolmogorov's $4/5$-law~(\ref{K45}). 
The distribution of vorticity is characterized by strongly non-Rayleigh shape, also indicating intermittency, and the power-law ``heavy tail'' of this distribution hints to a non-trivial geometry of the pancake vorticity structures, as we explain in the paper.\\

%----------------------------------------------------------------------------
%----------------------------------------------------------------------------

\textbf{2.} We solve the incompressible 3D Euler equations (in the vorticity formulation) 
\begin{equation}
\partial\omegabold/\partial t = \mathrm{rot}\,(\mathbf{v}\times \omegabold),\quad
\mathbf{v} = \mathrm{rot}^{-1}\omegabold,
\label{Euler2}
\end{equation}
numerically in the periodic box $\mathbf{r} = (x,y,z)\in [-\pi ,\pi ]^{3}$ with the pseudo-spectral Runge-Kutta fourth-order method. 
We start from the initial conditions taken as a superposition of the shear flow 
\begin{equation}\label{IC}
\omegabold_{sh}(\mathbf{r}) = (\sin z, \cos z, 0),\quad |\omegabold_{sh}(\mathbf{r})|=1,
\end{equation}
representing the stationary solution of the Euler equations, and a random truncated (up to second harmonics) periodic perturbation. 
The inverse of the curl operator and all the spatial derivatives are calculated in the Fourier space. 
We use an adaptive anisotropic rectangular grid, which is uniform for each direction and adapted independently along each of the three coordinates; the adaption comes from the analysis of the Fourier spectrum of the vorticity. 
Time stepping is implemented via the CFL stability criterion with the Courant number $0.5$. 
We start with the cubic grid $128^3$, refine the grid until the total number of nodes reaches $2048^{3}$ ($1024^3$ for some simulations), then fix the grid and continue until the Fourier spectrum of the vorticity at $2K_{\max }^{(j)}/3$ exceeds $10^{-13}$ times its maximum value along any of the three directions. 
Here $K_{\max}^{(j)}=N_{j}/2$ are the maximal wavenumbers and $N_{j}$ are numbers of nodes along directions $j=x,y,z$. 
For more details, we refer the reader to~\cite{agafontsev2015,agafontsev2016development,agafontsev2017universal}, where it was verified that the accuracy within the simulation time interval is very high and the simulations of the Euler equations transformed to the so-called vortex lines representation produce the same vorticity field. 

For some simulations, we observe the gradual formation of the power-law interval in the energy spectrum $E_{k}\propto k^{-\alpha}$ at small and moderate wavenumbers starting from $k\ge 2$. 
The first harmonic $k=1$, where the initial conditions were concentrated, contains most of the total energy (up to 97\% at the final time) and does not belong to this interval. 
To exclude its influence on the velocity structure functions, we calculate the moments for the modified velocity $\tilde{\mathbf{v}}$ obtained from the original velocity by setting the nine harmonics $\mathbf{k}=(k_{x},k_{y},k_{z})$ with $k_{x,y,z}= -1,0,1$ to zero.

In contrast to the developed (stationary) turbulence case, for which the moments can be computed using averaging in time, see e.g.~\cite{ishihara2009study} and the references wherein, calculation of the moments for the (non-stationary) problem of turbulence onset takes much more computational resources. 
We perform this calculation in the following way. 
First, for a given radius $r$, we find a sufficient number of points $\mathbf{r}$ evenly distributed on the sphere $|\mathbf{r}|=r$. 
Then, for each $\mathbf{r}$, we calculate the velocity variation $\mathbf{\delta \tilde{v}} = \tilde{\mathbf{v}}(\mathbf{x}+\mathbf{r},t)-\tilde{\mathbf{v}}(\mathbf{x},t)$ at every grid node $\mathbf{x}$, using the nearest-neighbor interpolation for the shifted velocity $\tilde{\mathbf{v}}(\mathbf{x}+\mathbf{r},t)$. 
Finally, the longitudinal and the transversal moments of order $n$, 
\begin{eqnarray}
M_{\parallel}^{(n)}(r) &=& \frac{1}{4\pi r^{2}} \int_{|\mathbf{r}|=r}d^{3}\mathbf{r}\int\frac{d^{3}\mathbf{x}}{(2\pi)^{3}}
\,(\mathbf{\delta \tilde{v}}\cdot \mathbf{m}_{r})^{n},  \label{moments-parallel}\\
M_{\perp}^{(n)}(r) &=& \frac{1}{4\pi r^{2}} \int_{|\mathbf{r}|=r}d^{3}\mathbf{r}\int\frac{d^{3}\mathbf{x}}{(2\pi)^{3}}
\,\bigg|\mathbf{\delta \tilde{v}}\times \mathbf{m}_{r}\bigg|^{n},  \label{moments-perpendicular}
\end{eqnarray}
where $\mathbf{m}_{r}=\mathbf{r}/r$ is the unit vector, are calculated as the corresponding integral sums over all points on the sphere $\mathbf{r}$ and all nodes $\mathbf{x}$.\\

%----------------------------------------------------------------------------
%----------------------------------------------------------------------------

\textbf{3.} We start with the simulation of the initial flow $I_{1}$ from~\cite{agafontsev2015} in grid limited by $2048^{3}$ total number of nodes; some details of this simulation were published previously in our paper~\cite{agafontsev2016asymptotic}. 
The simulation reaches the final time $t=7.75$ with the grid $972\times 2048\times 4096$ and the vorticity maximum $\omega_{\max}$ increased from $1.5$ to $18.4$, with the thinnest high-vorticity structure resolved with $10$ grid points at the level of vorticity half-maximum. 

\begin{figure}[t]
\centering
\includegraphics[width=7.5cm]{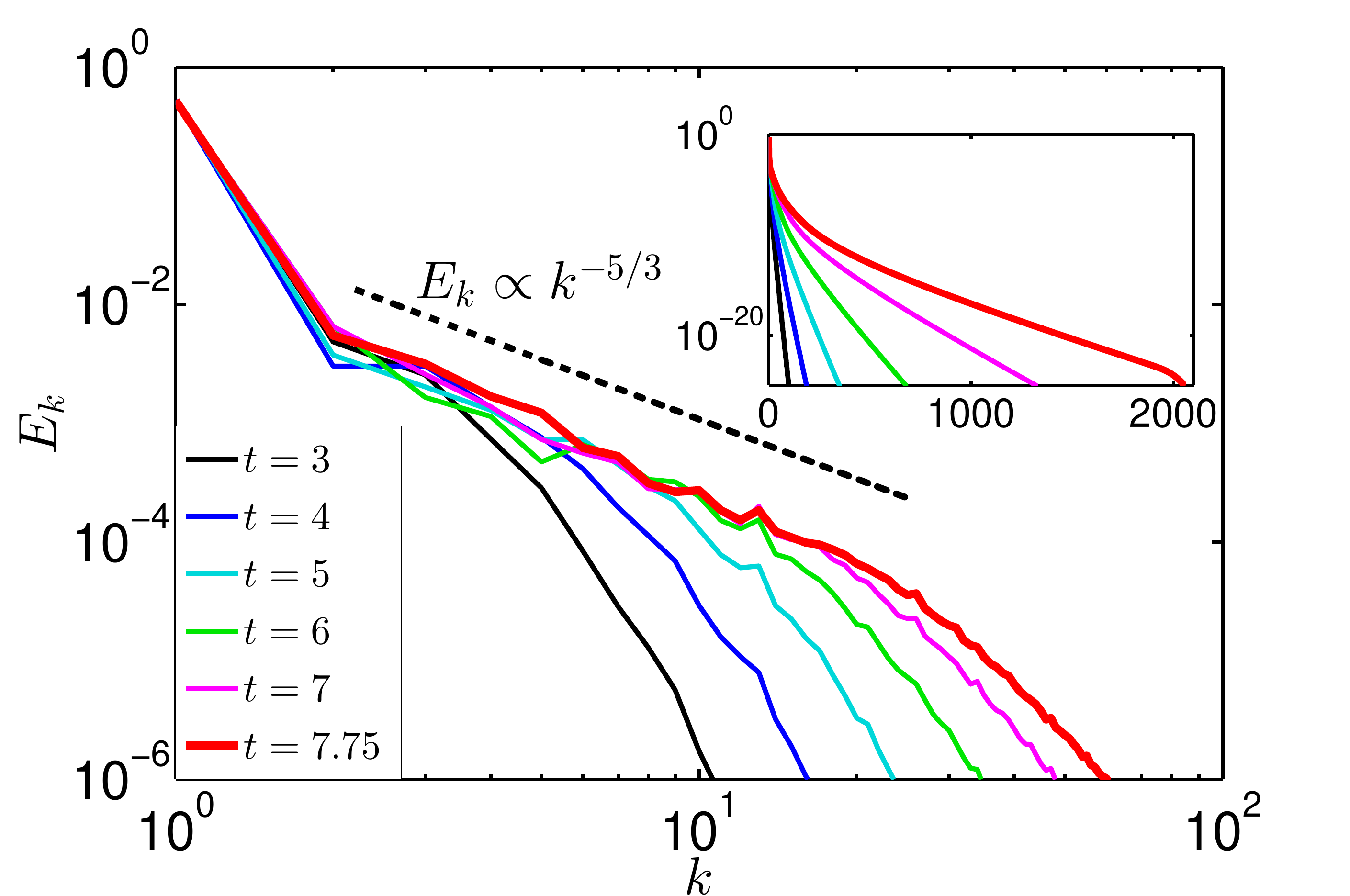}
\caption{{\it (Color on-line)} 
Energy spectrum $E_{k}$ in double-logarithmic scales, at different times. The inset shows the spectrum in semi-logarithmic scales.
}
\label{fig:fig1}
\end{figure}

The evolution of the energy spectrum for this simulation is shown in Fig.~\ref{fig:fig1}. 
At large wavenumbers $k$, the spectrum decays close to exponentially, as demonstrated by the inset of the figure. 
%The exponent $\delta(t)$ gradually decreases with time, that corresponds to fast excitation of small-scale motion; see~\cite{agafontsev2015} for more details on the behavior of $\delta(t)$. 
At small and moderate $k$, we clearly observe the gradual formation of the power-law interval with close to Kolmogorov's scaling $E_{k}\propto k^{-5/3}$. 
The power-law interval is characterized by the ``frozen'' spectrum, in contrast to the vast changes with time at larger wavenumbers, and extends up to a decade $2 \lesssim k \lesssim 30$ at the end of the simulation. 
Note that this interval acquires only a small fraction of the total energy: even at the final time, 97.2\% of energy is still contained in the first harmonic $k=1$, while wavenumbers $2\le k\le 30$ and $k>30$ obtain only 2.8\% and less than 0.1\% of energy, respectively. 

\begin{figure}[t]
\centering
\includegraphics[width=7.5cm]{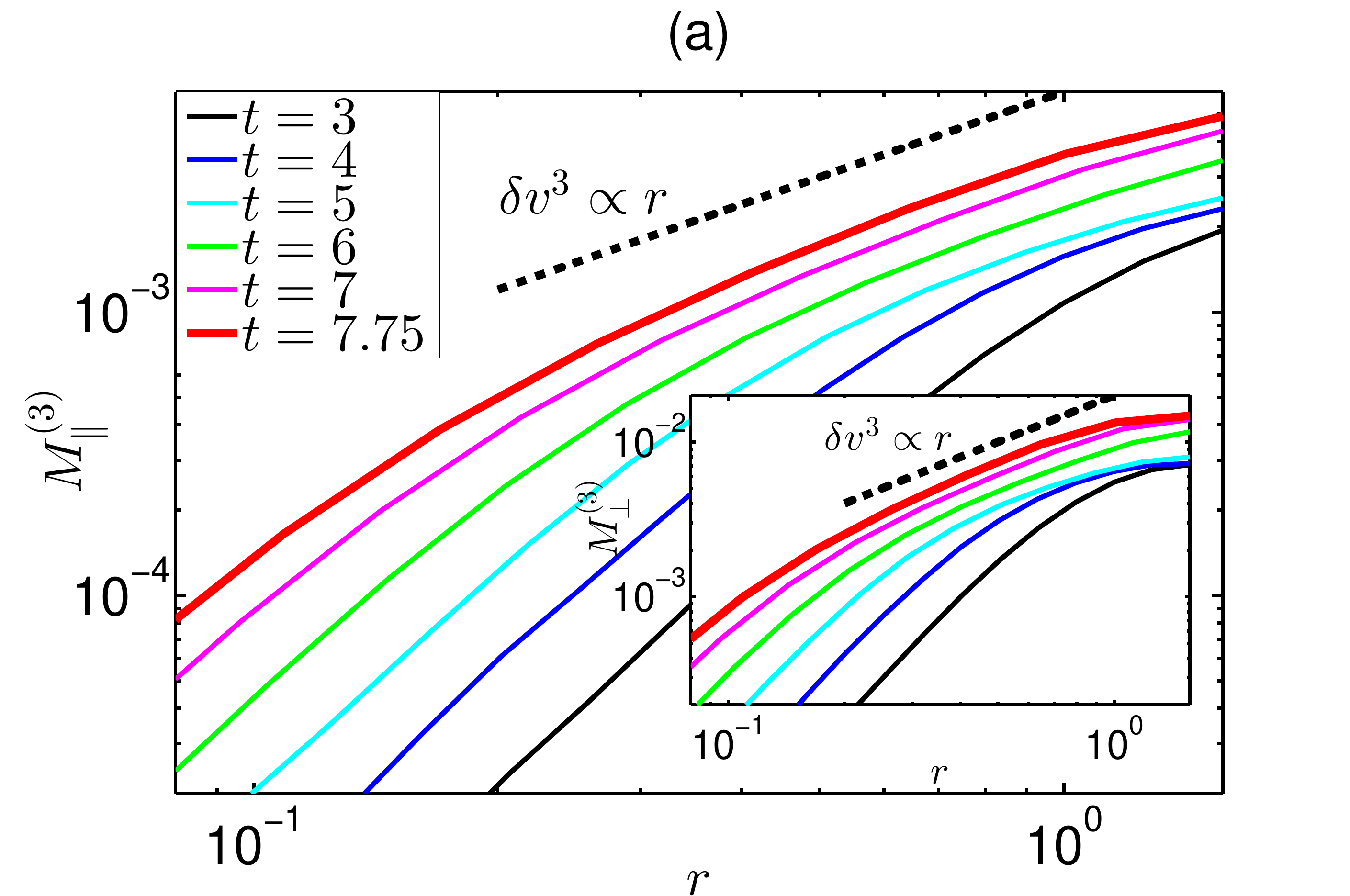}
\includegraphics[width=7.5cm]{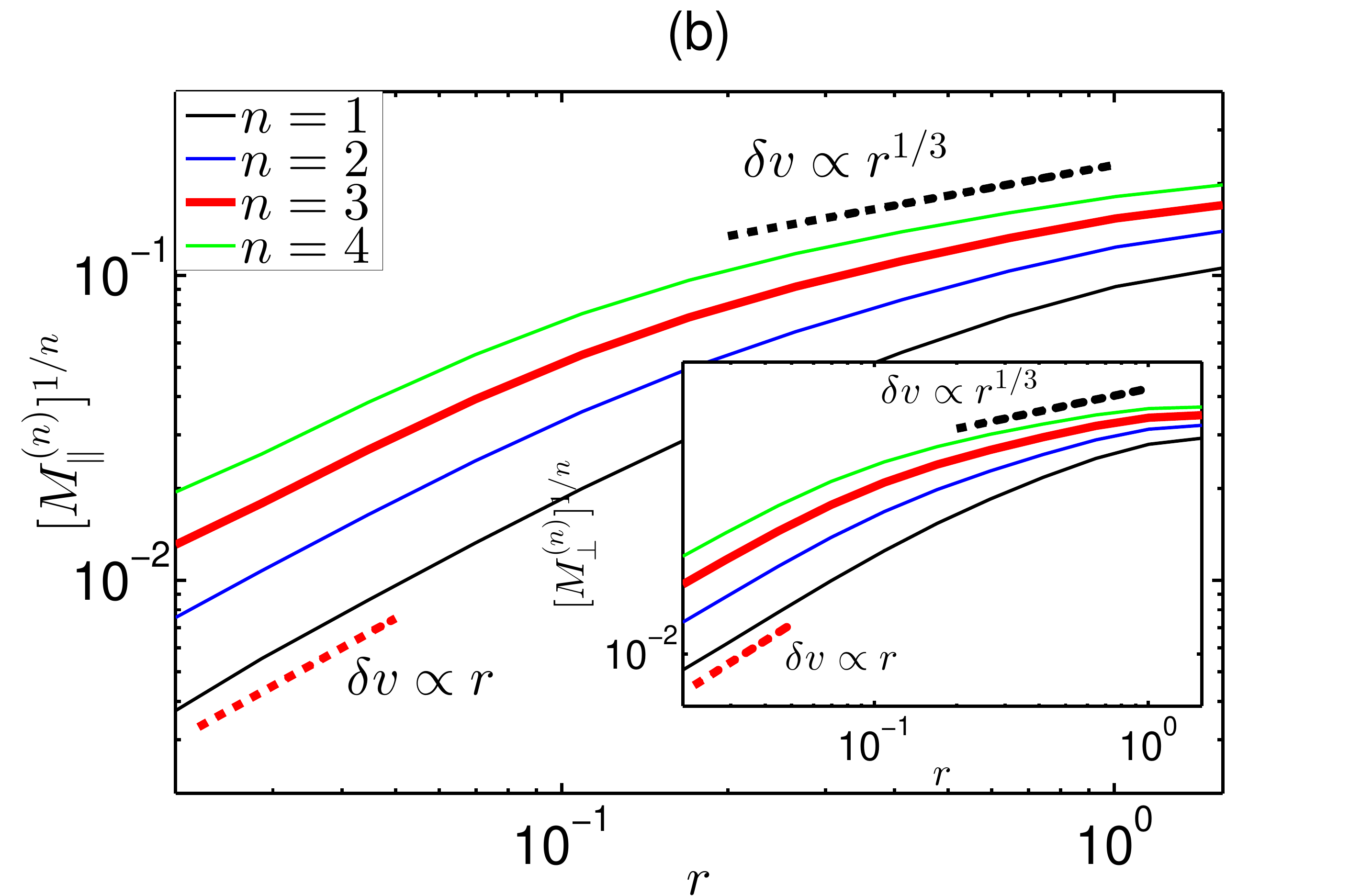}
\includegraphics[width=7.5cm]{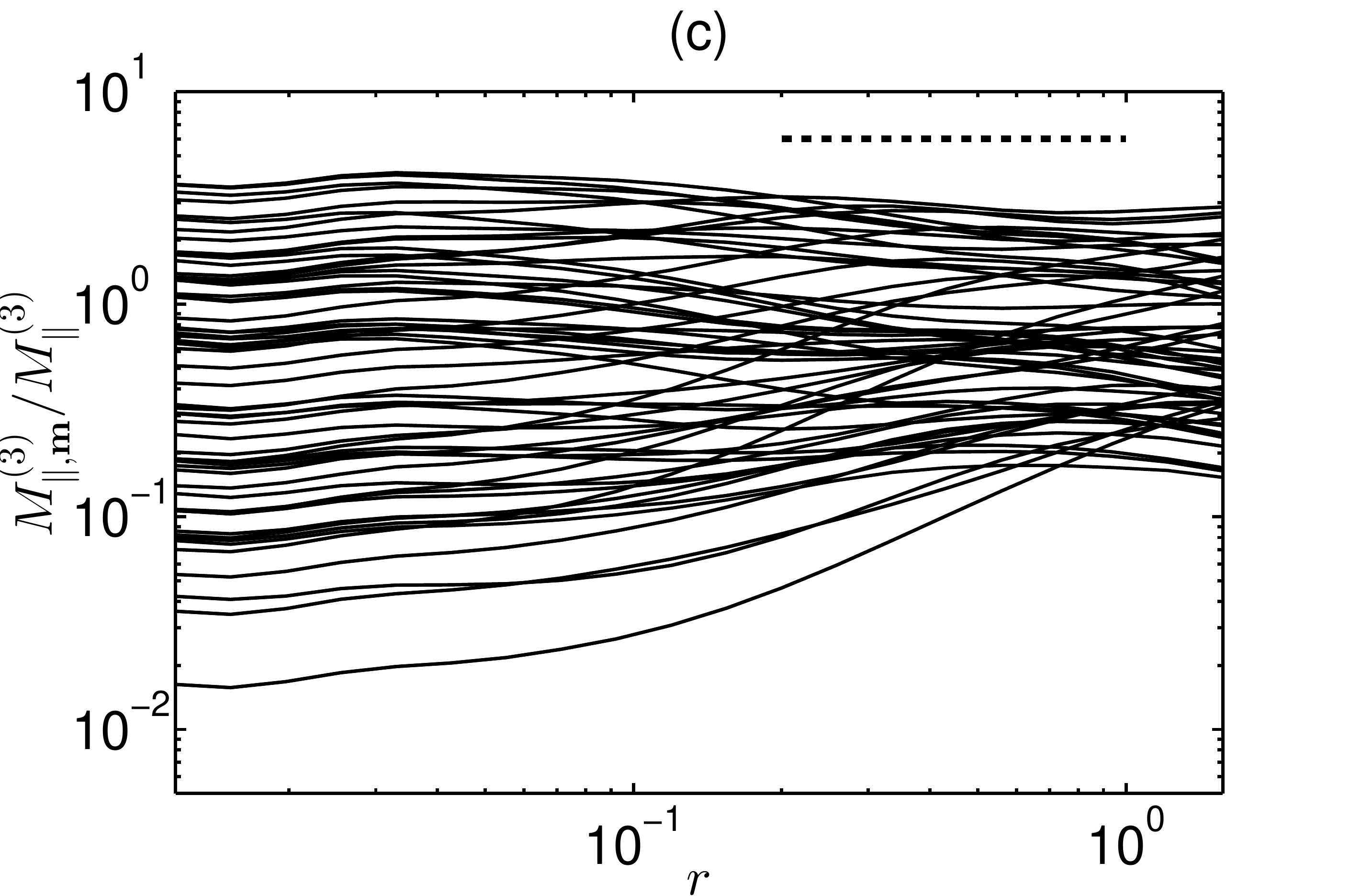}
\caption{{\it (Color on-line)} 
(a) Longitudinal third-order moments $M_{\parallel}^{(3)}$, at different times. The black dashed line indicates the scaling $M^{(3)}\propto r$. 
(b) Longitudinal moments $[M_{\parallel}^{(n)}]^{1/n}$ of orders $n=1,2,3,4$, at the final time $t=7.75$. The black dashed line indicates the scaling $[M^{(n)}]^{1/n}\propto r^{1/3}$ at the power-law interval, while the red dashed line - the scaling $[M^{(n)}]^{1/n}\propto r$ at smaller scales. 
The insets in figures (a) and (b) show the transversal moments. 
(c) Compensated directional longitudinal third-order moment $M_{\parallel,\mathbf{m}}^{(3)}/M_{\parallel}^{(3)}$ for $114$ directions $\mathbf{m}$ evenly distributed over the spherical coordinates, at the final time. 
The dashed horizontal line indicates the power-law interval for $M_{\parallel}^{(3)}$ in figure (b). 
}
\label{fig:fig2}
\end{figure}

The distribution of the velocity field linked to the power-law energy spectrum can be examined with the moments of velocity. 
Since the first harmonic containing most of the energy does not belong to the power-law interval, we exclude it from the analysis as explained above. 
The evolution of the third-order moments is shown in Fig.~\ref{fig:fig2}(a); see also Fig.~\ref{fig:fig2}(b) for the moments at the final time illustrated with larger scale. 
The power-law interval with close to linear scaling with the distance $M^{(3)}\propto r$ gradually forms for both the longitudinal and the transversal moments at sufficiently large scales, extending up to $0.2\lesssim r\lesssim 1$ at the final time. 
These scales correspond to wavenumbers $6\lesssim k\lesssim 30$ belonging to the power-law interval in the energy spectrum in Fig.~\ref{fig:fig1}. 

At the power-law interval, the scaling exponents $\xi_{n}$ and $\zeta_{n}$ for the longitudinal and the transversal moments $[M_{\parallel}^{(n)}]^{1/n}\propto r^{\xi_{n}}$ and $[M_{\perp}^{(n)}]^{1/n}\propto r^{\zeta_{n}}$ decrease with the order $n$, indicating both the intermittency and the anomalous scaling; see Fig.~\ref{fig:fig2}(b). 
The numerical values of the first four longitudinal exponents are $\xi_{1}=0.60 \pm 0.06$, $\xi_{2} = 0.48 \pm 0.04$, $\xi_{3} = 0.39 \pm 0.03$ and $\xi_{4} = 0.32 \pm 0.03$. 
The corresponding transversal exponents $\zeta_{1} = 0.55 \pm 0.07$, $\zeta_{2} = 0.42 \pm 0.06$, $\zeta_{3} = 0.33 \pm 0.05$ and $\zeta_{4} = 0.26 \pm 0.04$ are slightly smaller, $\xi_{n}\gtrsim\zeta_{n}$, but remain within the range of the standard deviations. 
Note that for the developed turbulence case the transversal exponents also turn out to be slightly smaller than the longitudinal ones, see e.g.~\cite{gotoh2002velocity,zybin2015stretching}. 

The anisotropy of the velocity distribution can be studied with the directional moments of velocity, for instance, the longitudinal third-order moment 
\begin{eqnarray}
M_{\parallel,\mathbf{m}}^{(n)}(r) &=& \int\frac{d^{3}\mathbf{x}}{(2\pi)^{3}}
\,(\mathbf{\delta \tilde{v}}\cdot \mathbf{m})^{n},  \label{moments-parallel-d}
\end{eqnarray}
where $\mathbf{r} = \mathbf{m}\,r$ and $\mathbf{m}$ is the unit vector setting the direction. 
The behavior of the directional moment $M_{\parallel,\mathbf{m}}^{(3)}$ relative to the angle-averaged one $M_{\parallel}^{(3)}$ is shown in Fig.~\ref{fig:fig2}(c) for $114$ directions evenly distributed over the spherical coordinates. 
At the scales of the power-law interval, the directional moment $M_{\parallel,\mathbf{m}}^{(3)}$ changes by up to order of magnitude with the direction, and for some directions it increases significantly faster (slower) with the distance $r$ than the angle-averaged moment $M_{\parallel}^{(3)}$. 
Note, however, that for many directions the directional moment changes with the distance very similarly to the angle-averaged one. 
Such behavior was first noted for two-dimensional hydrodynamic turbulence in the direct cascade regime~\cite{kuznetsov2015anisotropic}, where the Kraichnan spectrum arises due to the vorticity quasi-shocks~\cite{kuznetsov2007effects,kudryavtsev2013statistical} analogous to the pancake vorticity structures of the 3D case. 
For the transversal directional moments we observe the same behavior as discussed for the longitudinal ones. 

In order to examine the connection between the energy spectrum and the moments of velocity in more detail, we perform additional $30$ simulations in grids limited by $1024^{3}$ total number of nodes for $30$ initial flows taken as a superposition of the shear flow~(\ref{IC}) and a random periodic perturbation 
\begin{equation}
\omegabold_{p}(\mathbf{r}) = \sum_{\mathbf{h}} 
\left[\mathbf{A}_\mathbf{h}\cos(\mathbf{h}\cdot\mathbf{r})
+\mathbf{B}_\mathbf{h}\sin(\mathbf{h}\cdot\mathbf{r})\right]. \label{IC1}
\end{equation}
Here $\mathbf{h} = (h_x,h_y,h_z)$ is a vector of integer components $|h_{j}|\le 2$, $j=x,y,z$, while vectors $\mathbf{A}_\mathbf{h}$ and $\mathbf{B}_\mathbf{h}$ of real random coefficients with zero mean and variance $\sigma_{\mathbf{h}}^2 \sim \exp(-|\mathbf{h}|^2)$ satisfy the orthogonality conditions, $\mathbf{h}\cdot\mathbf{A}_\mathbf{h} = \mathbf{h}\cdot\mathbf{B}_\mathbf{h} = 0$, necessary for self-consistency. 
The initial conditions are taken as the mix of the flows~(\ref{IC}) and~(\ref{IC1}), 
\begin{equation}
\omegabold_{0}(\mathbf{r}) = (1-p)\,\omegabold_{sh}(\mathbf{r}) + p\, R\, \omegabold_{p}(\mathbf{r}),
\label{IC2}
\end{equation}
where $p$ is the mixing coefficient and $R=\sqrt{4\pi^{3}/E_{p}}$ is the renormalization coefficient. 
Here $4\pi^{3}$ and $E_{p}$ are the total energies of the shear flow~(\ref{IC}) and the perturbation~(\ref{IC1}) in the computational box $[-\pi ,\pi ]^{3}$, so that the coefficient $R$ renormalizes the perturbation to the same energy as has the shear flow. 
We perform three groups of experiments with $p=1$ (generic periodic flows), $p=0.1$ and $p=0.02$, with $10$ random realizations of initial flows for each group. 
Note that in~\cite{agafontsev2016development} we examined the similar flows, however, the initial conditions were constructed by adjusting the maximal vorticity of the perturbation, not its total energy as in this paper. 

\begin{figure}[t]
\centering
\includegraphics[width=7.5cm]{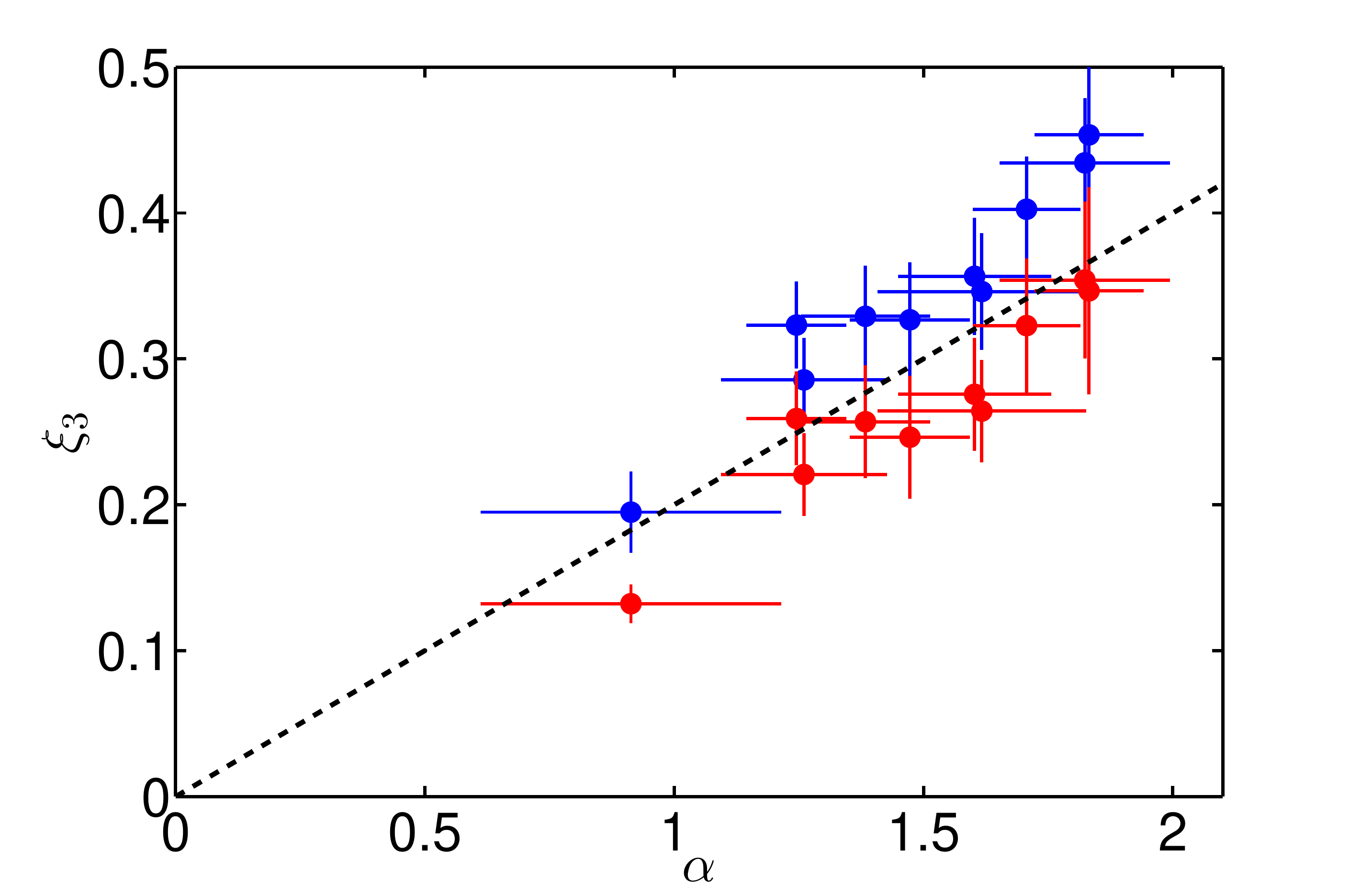}
\caption{{\it (Color on-line)} Exponents $\xi_{3}$ (blue) and $\zeta_{3}$ (red) for the power-law scalings of the longitudinal and the transversal third-order moments vs. exponent $\alpha$ for the power-law scaling of the energy spectrum; $10$ simulations of the third group of experiments with $p=0.02$. 
Horizontal and vertical lines indicate the standard deviations, the dashed black line shows the relation $\xi_{3}=\zeta_{3}=\alpha/5$. 
}
\label{fig:fig3}
\end{figure}

For the first group of experiments with generic periodic flows, none of the ten simulations develops power-law interval for the energy spectrum or for the moments of velocity. 
For the second group $p=0.1$, all ten simulations demonstrate power-law interval for the energy spectrum and six out of ten simulations develop power-law regions for the moments of velocity; the intervals extend up to $2\lesssim k\lesssim 20$ for the spectrum and $0.3\lesssim r\lesssim 0.8$ for the moments. 
The third group with $p=0.02$ shows power-law intervals for both the spectrum and the moments for all ten simulations; the intervals extend up to $2\lesssim k\lesssim 40$ and $0.15\lesssim r\lesssim 0.8$, respectively. 
For all simulations, the lower border $r_{l}$ of the power-law region $r_{l}\lesssim r\lesssim r_{h}$ for the moments (if this region is present) is related with the higher border $k_{h}$ of the power-law region $k_{l}\lesssim k\lesssim k_{k}$ for the spectrum as $r_{l}\approx 2\pi/k_{h}$. 
The higher border $r_{h}$ roughly corresponds to wavenumber $2\pi/r_{h} \simeq 6$.

For the third group of experiments, we observe the power-law scaling $E_{k}\propto k^{-\alpha}$ for the energy spectrum with the exponent $\alpha$ between $0.9$ and $1.8$, with most of the simulations having $\alpha$ close to $1.6$. 
The exponents $\xi_{3}$ and $\zeta_{3}$ describing the power-law scaling of the velocity moments $[M_{\parallel}^{(3)}(r)]^{1/3}\propto r^{\xi_{3}}$ and $[M_{\perp}^{(3)}(r)]^{1/3}\propto r^{\zeta_{3}}$ take values $0.2 \le\xi_{3}\le 0.45$ and $0.13 \le\zeta_{3}\le 0.35$. 
The longitudinal exponents turn out to be slightly larger than the transversal ones, $\xi_{3}\gtrsim\zeta_{3}$, and most of the ten simulations demonstrate $\xi_{3}$ close to $0.35$ and $\zeta_{3}$ close to $0.25$. 
As shown in Fig.~\ref{fig:fig3}, the simulations having larger exponent $\alpha$ also show larger exponents $\xi_{3}$ and $\zeta_{3}$, and vice versa, with the approximate relation for the longitudinal exponent
\begin{equation}
\xi_{3}\simeq \alpha/5. \label{zeta3-alpha}
\end{equation}
%The simulation of the $I_{1}$ initial flow discussed earlier has $\alpha\approx 5/3$, $\xi_{3}\approx 0.39$ and $\zeta_{3}\approx 0.33$, and also satisfies this relation. 
Note that this relation cannot be obtained from simple Fourier analysis. 
Indeed, a velocity variation satisfying $\delta v\propto r^{\zeta}$ in the physical space has the scaling $\delta v_{k}\propto k^{-\zeta-1}$ in the Fourier space, that leads to the energy spectrum $E_{k}\propto k^{-2\zeta-1}$. 
The two relations $\zeta = \alpha/5$ and $\zeta = (\alpha-1)/2$ intersect only at one point: $\alpha = 5/3$, $\zeta = 1/3$.

\begin{figure}[t]
\centering
\includegraphics[width=7.5cm]{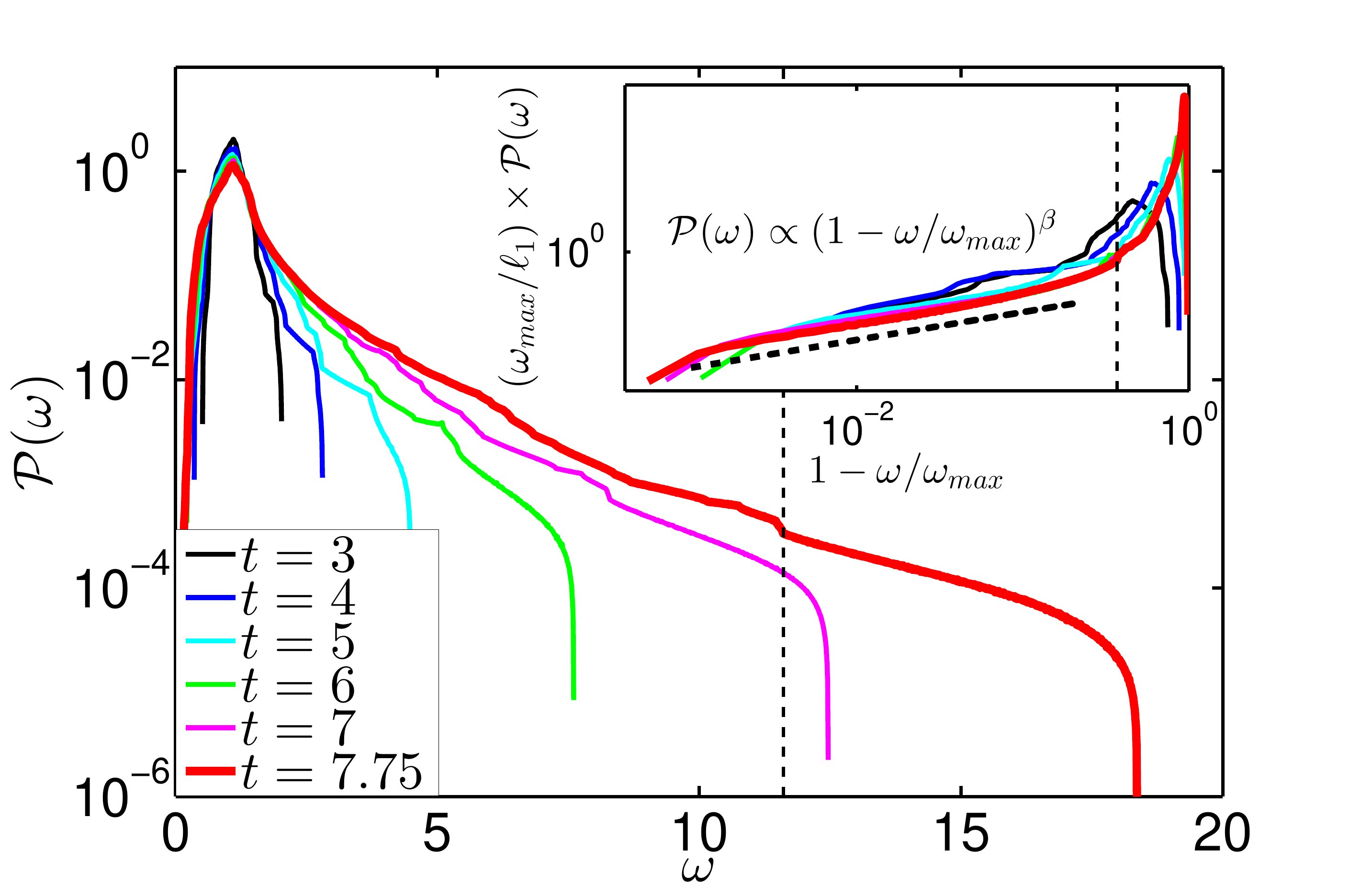}
\caption{{\it (Color on-line)} Distribution of vorticity for the simulation of the $I_{1}$ initial flow, at different times. 
The inset shows the normalized distribution of vorticity vs. $(1-\omega/\omega_{\max})$. 
The thin dashed vertical lines show the position of the second local vorticity maximum, while the thick dashed line in the inset indicates the scaling $\mathcal{P}(\omega)\propto (1-\omega/\omega_{\max})^{\beta}$ with $\beta=0.6$.
}
\label{fig:fig4}
\end{figure}

For all simulations with the power-law scaling of the moments, the scaling exponents $\xi_{n}$ and $\zeta_{n}$ decrease with the order $n$, indicating the intermittency. 
Another function that may hint to the intermittency is the distribution $\mathcal{P}(\omega)$ of the absolute value of vorticity. 
The evolution of this function for the $I_{1}$ simulation is shown in Fig.~\ref{fig:fig4}. 
The distribution shows strongly non-Rayleigh shape with the so-called ``heavy tail'' expanding to larger vorticity as the maximum vorticity increases with time. 
The value of the second local vorticity maximum (indicated in Fig.~\ref{fig:fig4} by the dashed vertical line) turns out to be significantly smaller than that of the first one, that allows us to study the vorticity distribution within the isolated pancake structure corresponding to the global vorticity maximum. 
In the local orthonormal basis $\mathbf{x} = \mathbf{x}_{m} + a_{1}\mathbf{w}_{1} + a_{2}\mathbf{w}_{2} + a_{3}\mathbf{w}_{3}$ of the pancake, the vorticity modulus can be described by the quadratic approximation~\cite{agafontsev2015}, 
\begin{equation}\label{pancake-quadratic}
\frac{|\omegabold(\mathbf{x})|}{\omega_{\max}} = 1 - \sum_{j=1}^{3}\bigg(\frac{a_{j}}{\ell_{j}}\bigg)^{2} + o(|\mathbf{x}-\mathbf{x}_{m}|^{2}),
\end{equation}
where $\mathbf{x}_{m}$ is the position of the vorticity maximum, $\ell_{j}=\sqrt{2\omega_{\max}/|\lambda_{j}|}$ are the characteristic pancake scales, $\ell_{1}\ll\ell_{2}\lesssim\ell_{3}$, while $\lambda_{1}<\lambda_{2}<\lambda_{3}<0$ and $\mathbf{w}_{j}$ are eigenvalues and unit eigenvectors of the (symmetric) Hessian matrix $\partial^{2}|\omegabold|/\partial x_{i}\partial x_{j}$ computed at $\mathbf{x}_{m}$. 
Using this approximation, we get 
$$
\mathcal{P}(f)\propto |dV/df| \propto (\ell_{1}\ell_{2}\ell_{3})(1-f)^{1/2},\quad f=\omega/\omega_{\max},
$$
where $V=(4\pi/3)\ell_{1}\ell_{2}\ell_{3}(1-f)^{3/2}$ is the volume of ellipsoid~(\ref{pancake-quadratic}). 
As we observed in~\cite{agafontsev2015}, only the pancake thickness $\ell_{1}$ significantly changes with time, while the other two scales $\ell_{2,3}$ remain of unity order. 
This allows to exclude $\ell_{2,3}$ from the above relation and leads to
\begin{equation}\label{PDF-scaling}
\mathcal{P}(\omega)\propto (\ell_{1}/\omega_{\max})(1-\omega/\omega_{\max})^{\beta},\quad \beta=1/2.
\end{equation}
Numerical simulations discussed in this paper are in good correspondence with the scaling~(\ref{PDF-scaling}), demonstrating though slightly larger value for the exponent $\beta$ between $0.5$ and $0.7$, see for example the inset in Fig.~\ref{fig:fig4} for the $I_{1}$ simulation. 
In our opinion, this discrepancy may reflect a non-trivial geometry of the pancake, which may deviate from its mid-plane much larger than the pancake thickness, see~\cite{agafontsev2015,agafontsev2016asymptotic}.\\

%----------------------------------------------------------------------------
%----------------------------------------------------------------------------

\textbf{4.} In conclusion, we have systematically examined the two-points structure functions (moments) of velocity. 
Despite the strong anisotropy inherent for the (non-stationary) problem of 3D hydrodynamic turbulence onset, we have observed formation of the power-law scaling for both the longitudinal and the transversal moments in the same interval of scales as for the energy spectrum. 
The scaling exponents for the velocity structure functions show the same key properties as for the developed (stationary) turbulence case. 
In particular, the exponents depend on the order of the moment non-trivially, indicating both the intermittency and the anomalous scaling, and the longitudinal exponents turn out to be slightly larger than the transversal ones. 
Analyzing several simulations for different initial conditions, we have arrived to a rough estimate $\xi_{3}\simeq \alpha/5$ between the scaling exponents for the longitudinal third-order moment and the energy spectrum. 
Thus, when the energy spectrum has power-law scaling close to the Kolmogorov's one, the longitudinal third-order moment shows close to linear scaling with the distance, in line with the Kolmogorov's $4/5$-law~(\ref{K45}). 
Note that before averaging over angles, the third-order moments demonstrate very anisotropic behavior, even though the linear scaling, as obtained after angle-averaging, can be traced back to most of the directions. 
The distribution of vorticity is characterized by strongly non-Rayleigh shape, also indicating the intermittency. 
The power-law scaling~(\ref{PDF-scaling}) for the tail of this distribution shows exponent $\beta\gtrsim 1/2$, that hints to a non-trivial geometry of the pancake vorticity structures. 

\textit{Acknowledgments}. The work of D.S.A. and E.A.K was supported by the Russian Science Foundation (grant 19-72-30028). 
The simulations were performed at the Novosibirsk Supercomputer Center (NSU), while the analysis of the results was done at the Data Center of IMPA (Rio de Janeiro). 
D.S.A. acknowledges the support from IMPA during the visits to Brazil. 
A.A.M. was supported by the RFBR (grant 17-01-00622) and the CNPq (grant 303047/2018-6).

%----------------------------------------------------------------------------
%----------------------------------------------------------------------------


\begin{thebibliography}{18}
\makeatletter
\providecommand \@ifxundefined [1]{%
 \@ifx{#1\undefined}
}%
\providecommand \@ifnum [1]{%
 \ifnum #1\expandafter \@firstoftwo
 \else \expandafter \@secondoftwo
 \fi
}%
\providecommand \@ifx [1]{%
 \ifx #1\expandafter \@firstoftwo
 \else \expandafter \@secondoftwo
 \fi
}%
\providecommand \natexlab [1]{#1}%
\providecommand \enquote  [1]{``#1''}%
\providecommand \bibnamefont  [1]{#1}%
\providecommand \bibfnamefont [1]{#1}%
\providecommand \citenamefont [1]{#1}%
\providecommand \href@noop [0]{\@secondoftwo}%
\providecommand \href [0]{\begingroup \@sanitize@url \@href}%
\providecommand \@href[1]{\@@startlink{#1}\@@href}%
\providecommand \@@href[1]{\endgroup#1\@@endlink}%
\providecommand \@sanitize@url [0]{\catcode `\\12\catcode `\$12\catcode
  `\&12\catcode `\#12\catcode `\^12\catcode `\_12\catcode `\%12\relax}%
\providecommand \@@startlink[1]{}%
\providecommand \@@endlink[0]{}%
\providecommand \url  [0]{\begingroup\@sanitize@url \@url }%
\providecommand \@url [1]{\endgroup\@href {#1}{\urlprefix }}%
\providecommand \urlprefix  [0]{URL }%
\providecommand \Eprint [0]{\href }%
\providecommand \doibase [0]{http://dx.doi.org/}%
\providecommand \selectlanguage [0]{\@gobble}%
\providecommand \bibinfo  [0]{\@secondoftwo}%
\providecommand \bibfield  [0]{\@secondoftwo}%
\providecommand \translation [1]{[#1]}%
\providecommand \BibitemOpen [0]{}%
\providecommand \bibitemStop [0]{}%
\providecommand \bibitemNoStop [0]{.\EOS\space}%
\providecommand \EOS [0]{\spacefactor3000\relax}%
\providecommand \BibitemShut  [1]{\csname bibitem#1\endcsname}%
\let\auto@bib@innerbib\@empty
%</preamble>
\bibitem [{\citenamefont {Kolmogorov}(1941)}]{kolmogorov1941dissipation}%
  \BibitemOpen
  \bibfield  {author} {\bibinfo {author} {\bibfnamefont {A.~N.}\ \bibnamefont
  {Kolmogorov}},\ }\href@noop {} {\bibfield  {journal} {\bibinfo  {journal}
  {Dokl. Akad. Nauk SSSR}\ }\textbf {\bibinfo {volume} {32}},\ \bibinfo {pages}
  {16} (\bibinfo {year} {1941})}\BibitemShut {NoStop}%
\bibitem [{\citenamefont {Landau}\ and\ \citenamefont
  {Lifshitz}(2013)}]{landau2013fluid}%
  \BibitemOpen
  \bibfield  {author} {\bibinfo {author} {\bibfnamefont {L.~D.}\ \bibnamefont
  {Landau}}\ and\ \bibinfo {author} {\bibfnamefont {E.~M.}\ \bibnamefont
  {Lifshitz}},\ }\href@noop {} {\emph {\bibinfo {title} {{Fluid Mechanics.
  Course of Theoretical Physics}}}},\ Vol.~\bibinfo {volume} {6}\ (\bibinfo
  {publisher} {Elsevier},\ \bibinfo {year} {2013})\BibitemShut {NoStop}%
\bibitem [{\citenamefont {Frisch}(1999)}]{frisch1999turbulence}%
  \BibitemOpen
  \bibfield  {author} {\bibinfo {author} {\bibfnamefont {U.}~\bibnamefont
  {Frisch}},\ }\href@noop {} {\emph {\bibinfo {title} {{Turbulence: the legacy
  of A.N.~Kolmogorov}}}}\ (\bibinfo  {publisher} {Cambridge University Press},\
  \bibinfo {year} {1999})\BibitemShut {NoStop}%
\bibitem [{\citenamefont {Orlandi}\ and\ \citenamefont
  {Pirozzoli}(2010)}]{orlandipirozzoli2010}%
  \BibitemOpen
  \bibfield  {author} {\bibinfo {author} {\bibfnamefont {P.}~\bibnamefont
  {Orlandi}}\ and\ \bibinfo {author} {\bibfnamefont {S.}~\bibnamefont
  {Pirozzoli}},\ }\href@noop {} {\bibfield  {journal} {\bibinfo  {journal}
  {Theor. Comput. Fluid Dyn.}\ }\textbf {\bibinfo {volume} {24}},\ \bibinfo
  {pages} {247} (\bibinfo {year} {2010})}\BibitemShut {NoStop}%
\bibitem [{\citenamefont {Holm}\ and\ \citenamefont
  {Kerr}(2002)}]{holm2002transient}%
  \BibitemOpen
  \bibfield  {author} {\bibinfo {author} {\bibfnamefont {D.~D.}\ \bibnamefont
  {Holm}}\ and\ \bibinfo {author} {\bibfnamefont {R.~M.}\ \bibnamefont
  {Kerr}},\ }\href@noop {} {\bibfield  {journal} {\bibinfo  {journal} {Phys.
  Rev. Lett.}\ }\textbf {\bibinfo {volume} {88}},\ \bibinfo {pages} {244501}
  (\bibinfo {year} {2002})}\BibitemShut {NoStop}%
\bibitem [{\citenamefont {Cichowlas}\ \emph {et~al.}(2005)\citenamefont
  {Cichowlas}, \citenamefont {Bona{\"\i}ti}, \citenamefont {Debbasch},\ and\
  \citenamefont {Brachet}}]{cichowlas2005effective}%
  \BibitemOpen
  \bibfield  {author} {\bibinfo {author} {\bibfnamefont {C.}~\bibnamefont
  {Cichowlas}}, \bibinfo {author} {\bibfnamefont {P.}~\bibnamefont
  {Bona{\"\i}ti}}, \bibinfo {author} {\bibfnamefont {F.}~\bibnamefont
  {Debbasch}}, \ and\ \bibinfo {author} {\bibfnamefont {M.}~\bibnamefont
  {Brachet}},\ }\href@noop {} {\bibfield  {journal} {\bibinfo  {journal} {Phys.
  Rev. Lett.}\ }\textbf {\bibinfo {volume} {95}},\ \bibinfo {pages} {264502}
  (\bibinfo {year} {2005})}\BibitemShut {NoStop}%
\bibitem [{\citenamefont {Holm}\ and\ \citenamefont {Kerr}(2007)}]{holm2007}%
  \BibitemOpen
  \bibfield  {author} {\bibinfo {author} {\bibfnamefont {D.~D.}\ \bibnamefont
  {Holm}}\ and\ \bibinfo {author} {\bibfnamefont {R.~M.}\ \bibnamefont
  {Kerr}},\ }\href@noop {} {\bibfield  {journal} {\bibinfo  {journal} {Phys.
  Fluids}\ }\textbf {\bibinfo {volume} {19}},\ \bibinfo {pages} {025101}
  (\bibinfo {year} {2007})}\BibitemShut {NoStop}%
\bibitem [{\citenamefont {Agafontsev}\ \emph {et~al.}(2015)\citenamefont
  {Agafontsev}, \citenamefont {Kuznetsov},\ and\ \citenamefont
  {Mailybaev}}]{agafontsev2015}%
  \BibitemOpen
  \bibfield  {author} {\bibinfo {author} {\bibfnamefont {D.~S.}\ \bibnamefont
  {Agafontsev}}, \bibinfo {author} {\bibfnamefont {E.~A.}\ \bibnamefont
  {Kuznetsov}}, \ and\ \bibinfo {author} {\bibfnamefont {A.~A.}\ \bibnamefont
  {Mailybaev}},\ }\href@noop {} {\bibfield  {journal} {\bibinfo  {journal}
  {Phys. Fluids}\ }\textbf {\bibinfo {volume} {27}},\ \bibinfo {pages} {085102}
  (\bibinfo {year} {2015})}\BibitemShut {NoStop}%
\bibitem [{\citenamefont {Agafontsev}\ \emph {et~al.}(2016)\citenamefont
  {Agafontsev}, \citenamefont {Kuznetsov},\ and\ \citenamefont
  {Mailybaev}}]{agafontsev2016development}%
  \BibitemOpen
  \bibfield  {author} {\bibinfo {author} {\bibfnamefont {D.~S.}\ \bibnamefont
  {Agafontsev}}, \bibinfo {author} {\bibfnamefont {E.~A.}\ \bibnamefont
  {Kuznetsov}}, \ and\ \bibinfo {author} {\bibfnamefont {A.~A.}\ \bibnamefont
  {Mailybaev}},\ }\href@noop {} {\bibfield  {journal} {\bibinfo  {journal}
  {JETP letters}\ }\textbf {\bibinfo {volume} {104}},\ \bibinfo {pages} {775}
  (\bibinfo {year} {2016})}\BibitemShut {NoStop}%
\bibitem [{\citenamefont {Brachet}\ \emph {et~al.}(1992)\citenamefont
  {Brachet}, \citenamefont {Meneguzzi}, \citenamefont {Vincent}, \citenamefont
  {Politano},\ and\ \citenamefont {Sulem}}]{brachet1992numerical}%
  \BibitemOpen
  \bibfield  {author} {\bibinfo {author} {\bibfnamefont {M.~E.}\ \bibnamefont
  {Brachet}}, \bibinfo {author} {\bibfnamefont {M.}~\bibnamefont {Meneguzzi}},
  \bibinfo {author} {\bibfnamefont {A.}~\bibnamefont {Vincent}}, \bibinfo
  {author} {\bibfnamefont {H.}~\bibnamefont {Politano}}, \ and\ \bibinfo
  {author} {\bibfnamefont {P.~L.}\ \bibnamefont {Sulem}},\ }\href@noop {}
  {\bibfield  {journal} {\bibinfo  {journal} {Phys. Fluids A}\ }\textbf
  {\bibinfo {volume} {4}},\ \bibinfo {pages} {2845} (\bibinfo {year}
  {1992})}\BibitemShut {NoStop}%
\bibitem [{\citenamefont {Agafontsev}\ \emph {et~al.}(2017)\citenamefont
  {Agafontsev}, \citenamefont {Kuznetsov},\ and\ \citenamefont
  {Mailybaev}}]{agafontsev2016asymptotic}%
  \BibitemOpen
  \bibfield  {author} {\bibinfo {author} {\bibfnamefont {D.~S.}\ \bibnamefont
  {Agafontsev}}, \bibinfo {author} {\bibfnamefont {E.~A.}\ \bibnamefont
  {Kuznetsov}}, \ and\ \bibinfo {author} {\bibfnamefont {A.~A.}\ \bibnamefont
  {Mailybaev}},\ }\href@noop {} {\bibfield  {journal} {\bibinfo  {journal} {J.
  Fluid Mech.}\ }\textbf {\bibinfo {volume} {813}} (\bibinfo {year}
  {2017})}\BibitemShut {NoStop}%
\bibitem [{\citenamefont {Agafontsev}\ \emph {et~al.}(2018)\citenamefont
  {Agafontsev}, \citenamefont {Kuznetsov},\ and\ \citenamefont
  {Mailybaev}}]{agafontsev2017universal}%
  \BibitemOpen
  \bibfield  {author} {\bibinfo {author} {\bibfnamefont {D.~S.}\ \bibnamefont
  {Agafontsev}}, \bibinfo {author} {\bibfnamefont {E.~A.}\ \bibnamefont
  {Kuznetsov}}, \ and\ \bibinfo {author} {\bibfnamefont {A.~A.}\ \bibnamefont
  {Mailybaev}},\ }\href@noop {} {\bibfield  {journal} {\bibinfo  {journal}
  {Phys. Fluids}\ }\textbf {\bibinfo {volume} {30}},\ \bibinfo {pages} {095104}
  (\bibinfo {year} {2018})}\BibitemShut {NoStop}%
\bibitem [{\citenamefont {Ishihara}\ \emph {et~al.}(2009)\citenamefont
  {Ishihara}, \citenamefont {Gotoh},\ and\ \citenamefont
  {Kaneda}}]{ishihara2009study}%
  \BibitemOpen
  \bibfield  {author} {\bibinfo {author} {\bibfnamefont {T.}~\bibnamefont
  {Ishihara}}, \bibinfo {author} {\bibfnamefont {T.}~\bibnamefont {Gotoh}}, \
  and\ \bibinfo {author} {\bibfnamefont {Y.}~\bibnamefont {Kaneda}},\
  }\href@noop {} {\bibfield  {journal} {\bibinfo  {journal} {Annu. Rev. Fluid
  Mech.}\ }\textbf {\bibinfo {volume} {41}},\ \bibinfo {pages} {165} (\bibinfo
  {year} {2009})}\BibitemShut {NoStop}%
\bibitem [{\citenamefont {Gotoh}\ \emph {et~al.}(2002)\citenamefont {Gotoh},
  \citenamefont {Fukayama},\ and\ \citenamefont {Nakano}}]{gotoh2002velocity}%
  \BibitemOpen
  \bibfield  {author} {\bibinfo {author} {\bibfnamefont {T.}~\bibnamefont
  {Gotoh}}, \bibinfo {author} {\bibfnamefont {D.}~\bibnamefont {Fukayama}}, \
  and\ \bibinfo {author} {\bibfnamefont {T.}~\bibnamefont {Nakano}},\
  }\href@noop {} {\bibfield  {journal} {\bibinfo  {journal} {Phys. Fluids}\
  }\textbf {\bibinfo {volume} {14}},\ \bibinfo {pages} {1065} (\bibinfo {year}
  {2002})}\BibitemShut {NoStop}%
\bibitem [{\citenamefont {Zybin}\ and\ \citenamefont
  {Sirota}(2015)}]{zybin2015stretching}%
  \BibitemOpen
  \bibfield  {author} {\bibinfo {author} {\bibfnamefont {K.~P.}\ \bibnamefont
  {Zybin}}\ and\ \bibinfo {author} {\bibfnamefont {V.~A.}\ \bibnamefont
  {Sirota}},\ }\href@noop {} {\bibfield  {journal} {\bibinfo  {journal} {Phys.
  Usp.}\ }\textbf {\bibinfo {volume} {58}},\ \bibinfo {pages} {556–}
  (\bibinfo {year} {2015})}\BibitemShut {NoStop}%
\bibitem [{\citenamefont {Kuznetsov}\ and\ \citenamefont
  {Sereshchenko}(2015)}]{kuznetsov2015anisotropic}%
  \BibitemOpen
  \bibfield  {author} {\bibinfo {author} {\bibfnamefont {E.~A.}\ \bibnamefont
  {Kuznetsov}}\ and\ \bibinfo {author} {\bibfnamefont {E.~V.}\ \bibnamefont
  {Sereshchenko}},\ }\href@noop {} {\bibfield  {journal} {\bibinfo  {journal}
  {JETP Lett.}\ }\textbf {\bibinfo {volume} {102}},\ \bibinfo {pages} {760–}
  (\bibinfo {year} {2015})}\BibitemShut {NoStop}%
\bibitem [{\citenamefont {Kuznetsov}\ \emph {et~al.}(2007)\citenamefont
  {Kuznetsov}, \citenamefont {Naulin}, \citenamefont {Nielsen},\ and\
  \citenamefont {Rasmussen}}]{kuznetsov2007effects}%
  \BibitemOpen
  \bibfield  {author} {\bibinfo {author} {\bibfnamefont {E.~A.}\ \bibnamefont
  {Kuznetsov}}, \bibinfo {author} {\bibfnamefont {V.}~\bibnamefont {Naulin}},
  \bibinfo {author} {\bibfnamefont {A.~H.}\ \bibnamefont {Nielsen}}, \ and\
  \bibinfo {author} {\bibfnamefont {J.~J.}\ \bibnamefont {Rasmussen}},\
  }\href@noop {} {\bibfield  {journal} {\bibinfo  {journal} {Phys. Fluids}\
  }\textbf {\bibinfo {volume} {19}},\ \bibinfo {pages} {105110} (\bibinfo
  {year} {2007})}\BibitemShut {NoStop}%
\bibitem [{\citenamefont {Kudryavtsev}\ \emph {et~al.}(2013)\citenamefont
  {Kudryavtsev}, \citenamefont {Kuznetsov},\ and\ \citenamefont
  {Sereshchenko}}]{kudryavtsev2013statistical}%
  \BibitemOpen
  \bibfield  {author} {\bibinfo {author} {\bibfnamefont {A.~N.}\ \bibnamefont
  {Kudryavtsev}}, \bibinfo {author} {\bibfnamefont {E.~A.}\ \bibnamefont
  {Kuznetsov}}, \ and\ \bibinfo {author} {\bibfnamefont {E.~V.}\ \bibnamefont
  {Sereshchenko}},\ }\href@noop {} {\bibfield  {journal} {\bibinfo  {journal}
  {JETP Lett.}\ }\textbf {\bibinfo {volume} {96}},\ \bibinfo {pages} {699}
  (\bibinfo {year} {2013})}\BibitemShut {NoStop}%
\end{thebibliography}
\end{document}